%% file: main.tex
\documentclass[a4paper,runningheads]{llncs}
\usepackage[T1]{fontenc}
\usepackage{graphicx}
\usepackage{array}
\usepackage{tikz}
\usetikzlibrary{positioning, arrows.meta, calc, fit}
\usepackage{hyperref}
\usepackage{orcidlink}
\usepackage{bbding}
\usepackage{tabularx}
\usepackage{url}
\let\arxivUrlBreaksPlainA\UrlBreaks\let\arxivUrlSpecialsPlainA\UrlSpecials\edef\arxivUrlmuskipPlainA{\the\Urlmuskip}
\usepackage{xurl}
\let\arxivUrlBreaksPkgA\UrlBreaks\let\arxivUrlSpecialsPkgA\UrlSpecials\edef\arxivUrlmuskipPkgA{\the\Urlmuskip}
\usetikzlibrary{positioning, arrows.meta, fit}
\hypersetup{hypertexnames=false}%
\newcommand{\artifactdoi}{10.5281/zenodo.23000442}
\newcommand{\repourl}{https://github.com/groundnuty/icsoc-2026-replication-package}
\newcommand{\file}[1]{\texttt{#1}}
\newlength{\arxivkeepA}
\newlength{\arxivkeepB}
\newlength{\arxivkeepC}
\newlength{\arxivkeepD}
\newlength{\arxivkeepE}
\let\arxivkeepF\topfraction
\let\arxivkeepG\bottomfraction
\let\arxivkeepH\textfraction
\let\arxivkeepI\floatpagefraction
\edef\arxivkeepK{\the\value{topnumber}}
\edef\arxivkeepL{\the\value{totalnumber}}
\renewcommand{\topfraction}{0.95}
\renewcommand{\bottomfraction}{0.9}
\renewcommand{\textfraction}{0.05}
\renewcommand{\floatpagefraction}{0.85}
\hypersetup{hidelinks}
\begin{document}
\let\UrlBreaks\arxivUrlBreaksPlainA\let\UrlSpecials\arxivUrlSpecialsPlainA\Urlmuskip=\arxivUrlmuskipPlainA\relax
\title{When the Consumer Is an Agent: Execution-Grounded Service-Level Agreements}
\titlerunning{When the Consumer Is an Agent}
\author{Micha\l{} Orzechowski\inst{1,2,3}(\Envelope)\orcidlink{0000-0002-8558-1283}
\and Renata G. S\l{}ota\inst{1}\orcidlink{0000-0001-7424-9317}
\and Jacek Kitowski\inst{1,2}\orcidlink{0000-0003-3902-8310}}
\authorrunning{M. Orzechowski et al.}
\institute{AGH University of Krakow, Poland\\
\email{morzech@agh.edu.pl} \and ACK Cyfronet AGH, Krakow, Poland \and Sano Centre for Computational Medicine, Krakow, Poland}

\maketitle

\begin{abstract}
Large language model (LLM) agents increasingly act as service consumers. They select
operations and report whether a task is complete. In an asynchronous distributed service, an
accepted call or immediate state read does not establish that the required state became observable by
its deadline. We introduce execution-grounded completion contracts, which treat completion
as a service-level outcome. An independent control plane observes a
deadline-bound service-state predicate, admits completion, and directs recovery while the
agent performs the task. We evaluate placement and deletion in a live multi-provider data
service with heterogeneous models and an injected transfer delay. Across policies,
some tasks that missed the deadline were accepted and some that met it were rejected. Under
enforcement, accepted completions remained consistent with later observations; all
included healthy tasks completed, and unresolved faulted tasks terminated as unverified.
The results show that completion of these tasks can be governed by independent observation
of service state rather than by the agent's self-assessment.

\keywords{Service-level agreements \and LLM agents \and Runtime verification \and Eventual
consistency \and Service reliability \and Federated data services.}
\end{abstract}

\input{sections/01-introduction}
\input{sections/02-background}
\input{sections/03-contracts}
\input{sections/04-evaluation-design}
\input{sections/05-results}
\input{sections/06-discussion}
\input{sections/07-threats}
\input{sections/08-conclusion}

\paragraph{Replication package.}
The replication package contains the trial records, endpoint and generation settings, run
order, setup-eligibility and reset procedures, deterministic verdict and attribution code, and
table-generation scripts. Every reported number regenerates exactly from the released traces;
the live runs deliberately used a real shared service, whose load and convergence vary. It is
available at \href{https://doi.org/10.5281/zenodo.23000442}{doi:10.5281/zenodo.23000442}.

\begin{credits}
\subsubsection{\ackname}
This work is co-financed by the Government Office of the Slovak Republic within the EU NextGenerationEU, Recovery and Resilience Plan under the SPICE project (09I02-03-V01-00012), by the Polish Ministry of Education and Science under the programs (5399/DIGITAL/2023/2, MEiN/2023/DIR/3796), by the EU Horizon 2020 program (No 857533), and by the subvention assigned to AGH University of Krakow. We acknowledge ACK Cyfronet AGH (PGINT/2026/019551, PLG/\-2026/019435, Gaia/AIF/101314359, LAIF/AIF/101234208) for providing computational resources.
\end{credits}

\input{part1.bbl}
\clearpage
\setcounter{totalnumber}{\arxivkeepL}
\setcounter{topnumber}{\arxivkeepK}
\let\floatpagefraction\arxivkeepI
\let\textfraction\arxivkeepH
\let\bottomfraction\arxivkeepG
\let\topfraction\arxivkeepF
\setlength{\belowcaptionskip}{\arxivkeepE}
\setlength{\abovecaptionskip}{\arxivkeepD}
\setlength{\intextsep}{\arxivkeepC}
\setlength{\floatsep}{\arxivkeepB}
\setlength{\textfloatsep}{\arxivkeepA}
\hypersetup{hidelinks}
\setlength{\emergencystretch}{2em}
\makeatletter\def\input@path{{artifact/}}\makeatother
\setcounter{section}{0}\setcounter{figure}{0}\setcounter{table}{0}\setcounter{equation}{0}\setcounter{footnote}{0}%
\let\UrlBreaks\arxivUrlBreaksPkgA\let\UrlSpecials\arxivUrlSpecialsPkgA\Urlmuskip=\arxivUrlmuskipPkgA\relax
\title{Artifact for When the Consumer Is an Agent: Execution-Grounded Service-Level Agreements}
\titlerunning{Artifact: When the Consumer Is an Agent}

\author{Micha\l{} Orzechowski\inst{1,2,3}(\Envelope)\orcidlink{0000-0002-8558-1283}
\and Renata G. S\l{}ota\inst{1}\orcidlink{0000-0001-7424-9317}
\and Jacek Kitowski\inst{1,2}\orcidlink{0000-0003-3902-8310}}
\authorrunning{M. Orzechowski et al.}
\institute{AGH University of Krakow, Poland\\
\email{morzech@agh.edu.pl} \and ACK Cyfronet AGH, Krakow, Poland \and Sano Centre for Computational Medicine, Krakow, Poland}

\maketitle

\begin{abstract}
This paper describes the artifact of the ICSOC 2026 paper `When the Consumer Is an Agent:
Execution-Grounded Service-Level Agreements'. The study lets LLM agents perform data
management tasks in a live Onedata federation and decides task completion by independent
observation of service state up to a deadline. The artifact contains the per-trial
recordings of the study, standard-library Python code that regenerates the paper's
tables and in-text numbers from these recordings with one command, the
experiment harness with its configuration, an offline test suite, and a container that
runs both checks without network access. The storage stack, the tool server, and the
harness are open source, and Onedata is deployed on Kubernetes from public Helm charts, so
the experiments can also be run anew on another deployment and with other models.

\keywords{Replication package \and Service-level agreements \and LLM agents \and Onedata \and Kubernetes}
\end{abstract}

\section{Introduction}

The paper~\cite{artifact-orzechowski2026} studies completion contracts for LLM agents that consume a
service. Before an agent starts a task, a contract fixes the required service state and a
deadline; an independent verifier observes that state; and a gate admits the agent's
completion only when the state holds. The evaluation runs seven models on data placement
and deletion tasks in a live federated data service, with and without an injected network
delay. The artifact supports three uses: verifying the reported numbers from the recorded
trials, checking the experiment harness without any service or model, and running the
experiments anew on one's own deployment and models.

The storage stack, the tool server, and the harness are open source. The federated service is
Onedata~\cite{artifact-onedata} (Apache-2.0), in which a Onezone service federates storage sites
that each run a Oneprovider. Both services are deployed on Kubernetes, here the k3s
distribution~\cite{artifact-k3s} (Apache-2.0), from the public Onedata Helm charts~\cite{artifact-onedatacharts}
(MIT) and the public Onedata container images. Onedata also underpins EGI
DataHub~\cite{artifact-egidatahub}, a central Onezone instance whose
default Oneprovider is operated by ACK Cyfronet AGH. The
agents reach the service through the onedata-mcp tool server~\cite{artifact-onedatamcp} (MIT), which
is built on the Model Context Protocol~\cite{artifact-mcp2025} Python SDK~\cite{artifact-mcpsdk} (MIT). The
harness and the scoring code in this artifact are released under the MIT License.

\begin{figure}[tb]
\centering
\input{artifact/fig-setup-body}
\caption{Experimental setup. The Onedata services run on Kubernetes (k3s) and are installed from the public Helm charts; each Oneprovider serves POSIX storage. The harness host runs the agent loop, the onedata-mcp tool server, the verifier, and the gate.}\label{artifact-fig:setup}
\end{figure}
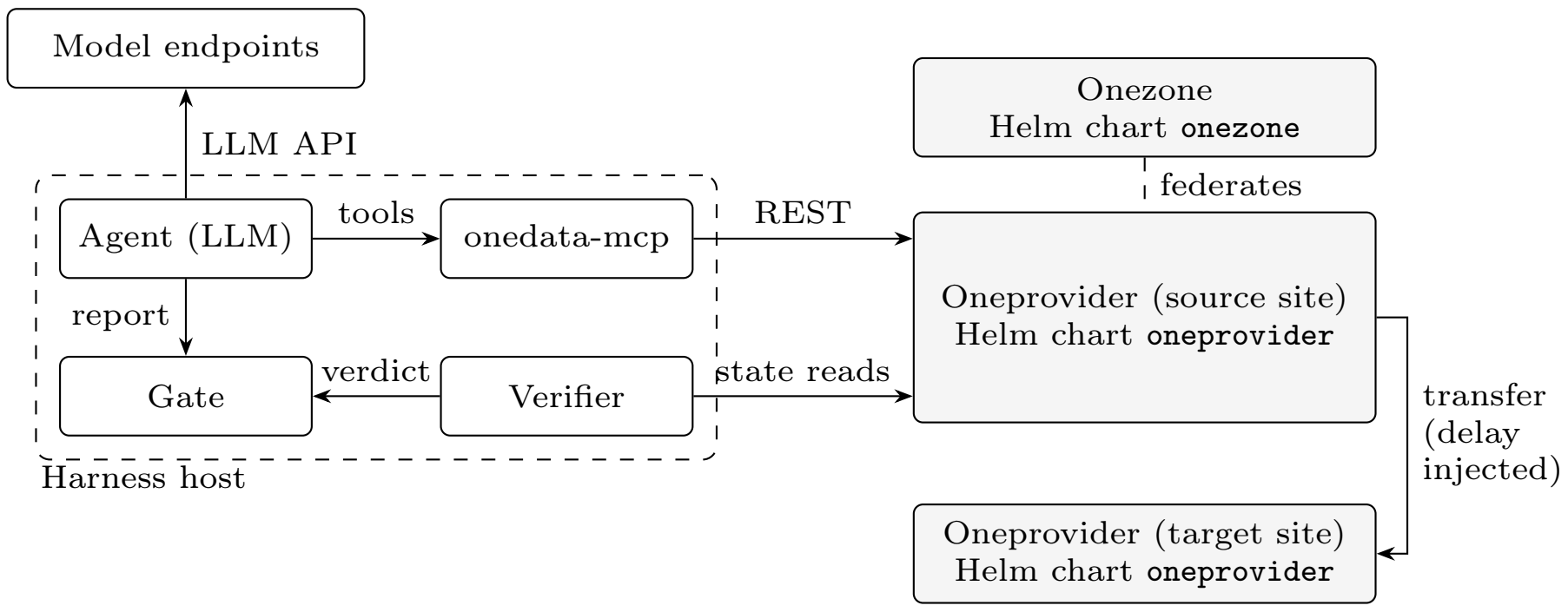

Five of the seven agent models are open-weight models (GLM-4.7-Flash, GLM-5.2-FP8,
Qwen3-Coder-30B, gemma-4-31B, and Llama-3.3-70B) served through an OpenAI-compatible
endpoint. The other two (claude-haiku-4-5 and claude-sonnet-5), the transcript judge, and
the contract-derivation pilot use Anthropic's commercial API through the Claude Code
command-line tool. Each model is one entry in a
configuration file, so other models can be substituted without code changes.
Figure~\ref{artifact-fig:setup} shows the experimental setup.

The artifact~\cite{artifact-artifact} is maintained at \url{\repourl}. Zenodo archives all its
versions under \href{https://doi.org/\artifactdoi}{doi:\artifactdoi}; this paper describes
release 1.0.0, archived under
\href{https://doi.org/10.5281/zenodo.23000443}{doi:10.5281/zenodo.23000443}.

\section{Artifact Contents}

Table~\ref{artifact-tab:contents} lists the repository. Table~\ref{artifact-tab:paths} summarises the three
ways to use it. The first two need no account, no network access, and no payment; only a
new run needs a Onedata deployment and access to models. Both run on an ordinary machine
without a GPU: the unpacked package takes about 36\,MB, the Python environment of the
offline check about 300\,MB, and the container image about 630\,MB.

\begin{table}[tb]
\centering
\caption{Contents of the artifact.}\label{artifact-tab:contents}
\scriptsize
\begin{tabularx}{\textwidth}{@{}>{\raggedright\arraybackslash}p{4.6cm}X@{}}
\hline
\textbf{Path} & \textbf{Content} \\
\hline
\file{recordings/} & 779 files: one JSON record per trial, grouped by study population, and the runs of the verifier-overhead benchmark \\
\file{SHA256SUMS}, \file{checksums.py} & checksums of every recording \\
\file{regenerate.py}, \file{a6\_table.py}, \file{in\_text.py}, \file{scoring/} & scoring code (Python standard library only) \\
\file{outputs/} & regenerated table bodies and in-text numbers \\
\file{harness/}, \file{scored\_sweep\_driver.py}, \file{a4\_sweep\_driver.py} & experiment harness and run drivers \\
\file{config/} & deployment settings, model arms, model and judge settings \\
\file{prompts/} & every prompt, verbatim \\
\file{Makefile}, \file{Dockerfile}, \file{requirements-harness.txt} & entry points, container, pinned packages \\
\file{README.md}, \file{RECORDINGS.md}, \file{ENVIRONMENT.md}, \file{VERIFICATION.md} & usage, record format, environment, transcript of a clean run \\
\hline
\end{tabularx}
\end{table}

\begin{table}[tb]
\centering
\caption{Three ways to use the artifact.}\label{artifact-tab:paths}
\scriptsize
\begin{tabularx}{\textwidth}{@{}>{\raggedright\arraybackslash}p{1.8cm}>{\raggedright\arraybackslash}X>{\raggedright\arraybackslash}X>{\raggedright\arraybackslash}p{1.6cm}@{}}
\hline
\textbf{Path} & \textbf{What it shows} & \textbf{Needs} & \textbf{Time} \\
\hline
Data only & every table and in-text number, regenerated from the recordings & Python~$\geq$3.10 (standard library) and \file{make} & seconds \\
Offline check & the harness test suite, including checks against the recorded arms, contracts, and prompts & Python 3.12--3.14 and pinned packages, or Docker; no network, no credentials & seconds after install \\
New run & a new run of the experiments & a Onedata deployment, model endpoints, API keys & hours \\
\hline
\end{tabularx}
\end{table}

\begin{table}[tb]
\centering
\caption{Computed tables of the paper and the files that regenerate them.}\label{artifact-tab:map}
\scriptsize
\begin{tabular}{@{}l l l@{}}
\hline
\textbf{Paper} & \textbf{Content} & \textbf{Output file} \\
\hline
Table 3 & study cells & \file{tab-study-matrix-body.tex} \\
Table 4 & read-back classes of false completion reports & \file{tab2-readback-classes-body.tex} \\
Table 5 & placement-policy error incidence & \file{tab3-placement-misjudgment-body.tex} \\
Table 6 & deadline sensitivity & \file{tab10-deadline-sensitivity-body.tex} \\
Table 7 & deletion false passes & \file{tab5-deletion-misjudgment-body.tex} \\
Table 8 & attribution accounting & \file{tab6-attribution-body.tex} \\
Table 9 & gate outcomes & \file{tab9-gate-outcomes-body.tex} \\
Table 10 & recovered first-attempt failures by mechanism & \file{tab7-rescue-mechanism-body.tex} \\
Table 11 & runtime-activity index & \file{tab8-cost-body.tex} \\
Table 12 & model charges with and without the gate & \file{tab11-cost-dollars-body.tex} \\
Sect.~5.3 & verifier overhead & \file{a6-overhead-table.md} \\
\hline
\end{tabular}
\end{table}

\section{Data-Only Verification}

From a clone of the repository or the unpacked Zenodo archive, run \texttt{make check}.
The command verifies the 779 recording files against \file{SHA256SUMS}, checks that
each study population holds exactly the expected files, regenerates every output, and
compares each output byte for byte with the committed copy. Its last line reads
\texttt{check: all 13 outputs byte-identical to the committed files}. It reads only
\file{recordings/}, uses no network and no model, and was tested with Python 3.10, 3.12,
3.13, and 3.14. On a mismatch it prints the difference and keeps the regenerated files in
\file{outputs.regenerated/}. \file{VERIFICATION.md} records the output of these commands and
of the offline check (Sect.~4), run on the release archive, for comparison.

Table~\ref{artifact-tab:map} maps the computed tables of the paper to their outputs. Every number
stated in the paper's text is in \file{outputs/in-text-numbers.txt}, one line per number
with its section, population, and level (trial, cycle, first attempt, or final). Two lines
differ from the printed text: the secondary adjudication in Sect.~5.2 of the paper is not
regenerated, because its data is not included, and for the admission-control count in
Sect.~5.4 the file gives the count over the included recordings. Tables~1 and~2 of the
paper are descriptive and are not computed.

\section{Offline Check}

The offline check runs the harness test suite without network access or credentials:
{\small
\begin{verbatim}
python3 -m venv .venv
.venv/bin/pip install -r requirements-harness.txt
make test PYTHON=.venv/bin/python
\end{verbatim}
}
\noindent or, in a container,
{\small
\begin{verbatim}
docker build -t consumer-agent-sla-artifact .
docker run --rm --network none consumer-agent-sla-artifact
\end{verbatim}
}
\noindent which runs both \texttt{make check} and \texttt{make test}. The suite has 442 tests,
and its output ends with \texttt{Ran 442 tests} and \texttt{OK}.
Besides unit tests of the harness, two tests tie the shipped code to the recordings: the
model arms built from \file{config/arms.json} have the recorded name, model settings, and
adapter for every one of the 448 recorded trials that use them, and the driver switches
rebuild the recorded contract and prompt byte for byte for all 168 deletion, informed, and
paired trials. The check exercises no live service and no model. It shows that the harness
runs and matches the recorded configuration; it does not show that a new run reproduces
the recorded behaviour.

\section{New Run on Your Own Deployment}

A new run needs a Onedata 25 deployment with one Onezone and two storage sites, a source and
a target, each running a Oneprovider with POSIX storage and reachable over HTTPS, and a
Onezone administrator credential. The services are deployed on any Kubernetes cluster from
the public Helm charts (Table~\ref{artifact-tab:env} lists the versions used), with site-specific
values such as domain names, TLS certificates, and the storage backend, and with the
image set to the 25.0 release:
{\scriptsize
\begin{verbatim}
helm repo add onedata https://onedata.github.io/charts
helm install onezone onedata/onezone --version 0.2.20-rc2 \
  --set image=onedata/onezone:25.0 -f onezone.yaml
helm install oneprovider onedata/oneprovider --version 0.2.20-rc7 \
  --set image=onedata/oneprovider:25.0 -f site.yaml
\end{verbatim}
}
\noindent with one Oneprovider release per storage site. The harness creates a fresh space for
each run, supports it on both sites, writes its own trial files, and removes the space at
the end. The agents use the onedata-mcp tool server at the commit in Table~\ref{artifact-tab:env}.
Model access is set in \file{config/arms.json}, with one entry per model arm: the adapter
kind (\texttt{openai-compatible} or \texttt{anthropic-sdk}), the model identifier, the base
URL, and the name of the environment variable that holds the API key. Runs with the injected
delay also need access to the target site's cluster with permission to attach a debug
container with \texttt{NET\_ADMIN} to the target Oneprovider pod. Deployment settings are
in \file{config/federation.json} (Sect.~6). A placement run without the fault starts with
{\small
\begin{verbatim}
RUN_ID=placement SWEEP_FAULT=none .venv/bin/python scored_sweep_driver.py
\end{verbatim}
}
\noindent and further environment variables select the other experiments:
{\scriptsize
\begin{verbatim}
SWEEP_FAULT=netem_lag      injected delay
SWEEP_SCENARIO=deletion    deletion task
SWEEP_INFORMED=1           prompt that also states the contract
\end{verbatim}
}
\noindent The gated variants run with \file{a4\_sweep\_driver.py}, and the overhead benchmark with
\file{harness/a6\_overhead.py}. New runs write records in the
format of \file{recordings/}. In the recorded runs, 56 trials without the fault took about
5--10 minutes, faulted trials about 1 minute each ungated and 2 minutes gated, and the
overhead benchmark about 80 minutes; the two Anthropic arms together cost USD 1.5--2.4 per
56-trial population. A new run is a new experiment, not a reproduction of the recorded one:
models are sampled, hosted models can be renamed or withdrawn, and a shared service has
its own timing.

\section{Extending the Artifact}

Another model is one more entry in \file{config/arms.json}, for example
{\scriptsize
\begin{verbatim}
{"name": "my-model", "kind": "openai-compatible", "model_id": "org/model",
 "base_url": "https://host/v1", "key_env": "MY_API_KEY"}
\end{verbatim}
}
\noindent after which \texttt{SWEEP\_LEGS=org/model} selects it and \texttt{SWEEP\_K} sets the
number of repetitions. An optional \texttt{api\_model\_id} sends a different identifier to
the endpoint while the record keeps \texttt{model\_id}. Another deployment is described in
\file{config/federation.json}: the Onezone URL; for each storage site its label, provider
identifier, host, and storage identifier; and, for the injected delay, the coordinates of
the target Oneprovider pod. Each key can also be set by an environment variable
\texttt{HARNESS\_<KEY>}.

\section{Data Origins and Provenance}

The recordings are the harness output of the study runs, stored as recorded and covered by
the checksums. Each trial record stores the requested model identifier and its generation
settings, the verbatim prompt, the agent's messages and tool calls with their results, the
verifier's state timeline, the gate outcome on gated runs, cost accounting, the start time,
and the harness version that produced it; \file{RECORDINGS.md} documents the format. The
files moved in the trials are 4096-byte files written by the harness, and \file{prompts/}
holds every prompt given to the agents, the judge, and the derivation pilot.

\section{Experiment Environment}

Table~\ref{artifact-tab:env} lists the environment of the recorded runs; \file{ENVIRONMENT.md} gives
the details.

\begin{table}[tb]
\centering
\caption{Environment of the recorded runs.}\label{artifact-tab:env}
\scriptsize
\begin{tabularx}{\textwidth}{@{}>{\raggedright\arraybackslash}p{2.6cm}X@{}}
\hline
\textbf{Component} & \textbf{Version or setting} \\
\hline
Onedata & 25.0 (Onezone and Oneprovider) \\
Kubernetes & k3s \texttt{v1.34.1+k3s1} \\
Helm charts & \url{https://onedata.github.io/charts}: \texttt{onezone 0.2.20-rc2}, \texttt{oneprovider 0.2.20-rc7} \\
Topology & one Onezone; two storage sites, each a Oneprovider with POSIX storage \\
Agent tools & onedata-mcp commit \texttt{7973925}; MCP Python SDK 1.28.1 (protocol 2025-11-25) \\
Trial parameters & deadline 30\,s; verifier polling every 3\,s up to 45\,s; injected delay 35\,s on the target's inter-site transfer traffic \\
Verifier host & AMD Opteron 62xx class CPU, 16 cores; Python 3.14.6 \\
\hline
\end{tabularx}
\end{table}

\section{License}

The code is released under the MIT License; the recordings and the prompts are released
under CC BY 4.0 (\file{LICENSE}). The licences of the components are given in Sect.~1.
Citation metadata are in \file{CITATION.cff}.

\input{part2.bbl}
\end{document}

%% file: sections/01-introduction.tex
\section{Introduction}\label{sec:introduction}

LLM agents increasingly act as service consumers on behalf of users: they interpret a goal,
select and invoke service operations, and report whether the requested task is complete.
Unlike a deterministic program executing a fixed call sequence, each decision can fail even
when every service call succeeds. An agent can therefore report completion despite contrary
service state~\cite{advani2026,rabanser2026}.

Asynchronous distributed services make that report difficult to assess. An accepted operation
may not yet have reached every component, and data and status signals may converge at
different times~\cite{vogels2009,decandia2007}. A report-time read can capture an intermediate
state rather than establish whether the user's goal held by its deadline. Operation acceptance
and task fulfillment are therefore distinct; completion is a service-level question about
whether the requested state became observable in time.

Existing mechanisms constrain agent actions and tool use, assess execution traces, or inspect
recorded environment state~\cite{wang2026,liu2026,chen2025,zheng2023,yao2025}. These
mechanisms improve execution-time observability and control but leave a service-level
requirement: an authoritative, deadline-aware completion decision for the requested state.

Service-level agreement methodology separates a measurable objective from its assessment and
the consequences of missing it; WS-Agreement also permits guarantee terms that bind the
consumer, not only the provider~\cite{andrieux2007,keller2003}. We bring this perspective to
agent-mediated service tasks by treating completion as a service-level outcome rather than an
agent-authored judgment. Our central idea is to separate task execution from completion
authority: the agent performs the task, while an independent assessment of service state
determines whether the requested state became observable within the required time. We realize
this separation through execution-grounded completion contracts;
Sect.~\ref{sec:contracts} develops their operational model.

We combine consumer-obligated guarantee terms with independent state verification at an
agentic service consumer's completion boundary, and give a first live evaluation of the resulting
service-control decisions on a federated data service. The paper makes four contributions:

\begin{enumerate}
\item A service-control model for agent-consumer completion contracts: state predicate,
deadline, verdicts, evidence boundary, gate, and recovery outcomes.\looseness=-1
\item A comparison of four common completion policies with a deadline-aware reference over
recorded ungated trials, showing acceptance-side and rejection-side disagreements.
\item An evidence-based attribution rule evaluated against an injected fault, with shadowed
cases reported separately from cases in which both sides leave observable evidence.
\item A live evaluation of contract enforcement on two task types and seven models, reporting
admitted completions, explicit unverified terminations, recovery mechanisms, runtime activity,
and dollar cost.
\end{enumerate}

%% file: sections/02-background.tex
\section{Background and Related Work}\label{sec:background}

Service-level agreements separate a promise from the mechanism that measures it. WSLA
specifies and monitors measurable service objectives. WS-Agreement represents guarantee
terms, obligated parties, and assessment states~\cite{keller2003,andrieux2007}.
WS-Agreement permits terms that obligate either the provider or the consumer, and its
assessment vocabulary includes \emph{Fulfilled}, \emph{Violated}, and
\emph{NotDetermined}. Consumer obligations are therefore part of the standard model; the
open design question is how to make one operational at an LLM agent's completion boundary.

Most operational service objectives concern provider behavior, such as availability,
latency, and error budgets~\cite{beyer2016,openslo}. AgentSLA likewise treats the agent as a
provider with quality obligations~\cite{jouneaux2025}; verifiable-SLA work protects provider
telemetry from self-reporting through trusted execution and cryptographic
evidence~\cite{castillo2026}. Our setting differs in role and granularity: the governed
component consumes rather than provides a service, and the objective concerns one task's
required state.

The backend adds uncertainty to this consumer-side obligation. In an eventually consistent
service, an accepted operation can become visible at different sites at different times, and
status metadata may lead or lag the data it describes~\cite{vogels2009,decandia2007}.
Consequently, neither a successful call nor one immediate read establishes that a task-level
condition held by its deadline. The observer must evaluate a pre-specified condition against
state sampled over a time window.

Research on service composition supplies much of the machinery for time-bounded observation.
Dynamo uses runtime rules for BPEL compositions~\cite{baresi2005}. Event-calculus monitors
check service requirements~\cite{mahbub2004}; Cremona implements WS-Agreement creation and
monitoring~\cite{ludwig2004}; and Leitner et al. predict and prevent SLA violations in
composite services~\cite{leitner2010}. These systems observe and act on service properties.
Our problem adds an LLM-controlled consumer whose completion report is itself the governed
decision.

Agent-verification systems apply related ideas during execution. AgentSpec checks
triggered safety rules at runtime, and ToolGate uses preconditions to gate a tool call and
postconditions to decide whether its result may update a trusted symbolic
state~\cite{wang2026,liu2026}. STRATUS organizes cloud-operations agents in a state machine
and enforces a no-regression specification during mitigation~\cite{chen2025}. Interwhen uses
self or external verifiers to check properties of a reasoning trace and steer generation at
test time~\cite{bhat2026}. These approaches primarily govern actions, intermediate state, or
reasoning while a task is being executed.

Moving verification to the completion decision changes what is governed. Verify-Gated
Completion presents the gate pattern: an agent proposes completion and a read-only verifier
admits it, failing closed on weak evidence~\cite{nguyen2026}; its stated scope excludes
task-level success, comparison with ungated baselines, and recovery effectiveness. We keep the
gate and change its evidence, deadline semantics and recovery, and compare task-level decisions
with four common completion policies across seven models.
Table~\ref{tab:positioning} compares these approaches by the evidence they check, their
decision point, deadline, attribution, and the action after a failed check.

\begin{table}[tb]
  \centering
  \caption{Comparison of agent-verification approaches and their verification semantics.}
  \label{tab:positioning}
  \scriptsize
  \input{figures/tab-positioning-body}
\end{table}

Common completion checks use less outcome evidence. An LLM judge reads a transcript rather
than backend state~\cite{zheng2023}. Tool-agent benchmarks may inspect final environment
state, but ordinarily after the interaction rather than as a live admission
decision~\cite{yao2025}. A report-time state read improves on transcript-only judging but
still samples one point in an asynchronous transition. Sect.~\ref{sec:design} scores these
information positions as completion policies on the same ungated records.\looseness=-1

Together, these lines of work supply the parts of a completion contract. Runtime verification
provides bounded observation~\cite{leucker2009}, design by contract supplies pre-specified
postconditions~\cite{meyer1992}, test oracles assess observed outcomes~\cite{barr2015}, and
workflow engines add timeouts and retries~\cite{temporal}. Consumer-driven contract tests
instead validate interactions before deployment~\cite{pact}. The model developed next
combines these elements at a service-consuming agent's live completion boundary and uses the
verdict evidence to direct recovery.\thisbottomragged

%% file: figures/tab-positioning-body.tex
\begin{tabular}{@{}>{\raggedright\arraybackslash}p{1.95cm}>{\raggedright\arraybackslash}p{2.3cm}>{\raggedright\arraybackslash}p{2.05cm}>{\raggedright\arraybackslash}p{1.25cm}>{\raggedright\arraybackslash}p{1.6cm}>{\raggedright\arraybackslash}p{2.5cm}@{}}
\hline
\textbf{Approach} & \textbf{Evidence checked} & \textbf{Decision point} & \textbf{Deadline} & \textbf{Attribution} (agent or service) & \textbf{Action after a failed check} \\
\hline
AgentSpec~\cite{wang2026} & The agent's trajectory and next action & On a trigger: state change, before an action, or agent finish & No & No & Stop, ask the user, self-examine, or run a corrective action \\[3pt]
ToolGate~\cite{liu2026} & Tool preconditions and results against the agent's trusted symbolic state & At each tool call & No & No & Block the call or discard its result \\[3pt]
STRATUS~\cite{chen2025} & Severity of the managed system's state & After each mitigation transaction & No & No & Undo the transaction \\[3pt]
Verify-Gated\newline Completion~\cite{nguyen2026} & Claim and evidence packets, common ground, and task state & When the agent proposes completion & No & No & Withhold completion; recovery is a separate path \\[3pt]
This paper & A pre-specified predicate over live backend state, observed outside the agent's session & At completion, from observations up to the deadline & Yes & Yes, from runtime evidence & Withhold completion; the deadline violation stays recorded; recovery routed by the attributed cause \\
\hline
\end{tabular}

%% file: sections/03-contracts.tex
\section{Per-Task Completion Contracts}\label{sec:contracts}

A user supplies a bounded goal, for example, \emph{``replicate this file to site B and
confirm that it arrived.''} Before the agent receives it, an operator---a person, controller,
or agent-management system---instantiates a contract from a task template. The agent does not
write the condition against which it will be assessed.

The contract binds five elements: a machine-checkable predicate over observable service state,
the task start and deadline, the obligated component, and a recovery policy. Predicate
instantiation fixes target identifiers and expected values; in this study, the obligated
component is the agent. In WS-Agreement terminology, the predicate and deadline form the
objective, the agent is its obligated party, and observations act as service
indicators~\cite{andrieux2007}.

Contract evaluation requires evidence independent of the obligated agent. The verifier is a
deterministic program outside the agent's session, with read-only access to the same class of
service state exposed by the agent's tools. Its probes and results do not enter the agent's
context. The agent's report or interpretation therefore cannot become verdict evidence.

Once the task starts, the verifier samples the state named by the predicate on a fixed grid
until the deadline. A \emph{Fulfilled} verdict means that an observation no later than the
deadline satisfied the predicate. A \emph{Violated} verdict means that readable observations
did not establish the predicate in time. A \emph{NotDetermined} verdict means that the
verifier lacked sufficient readable evidence to decide. A failed probe cannot count as
satisfaction.

The deadline belongs to the original obligation. If the predicate becomes true only
afterward, the original attempt remains Violated. Later observations may record convergence
and support recovery, but cannot revise that verdict. This distinction separates the service
outcome eventually reached from the promise made for a particular time window.

\label{sec:attribution-model}
The verdict establishes whether the obligation was met, but not why it was missed. After a
Violated verdict, the verifier therefore considers two evidence streams. Agent-side evidence
records whether a valid operation targeted the canonical object in time; service-side evidence
records call outcomes and the subsequent state trajectory. The implementation emits
\emph{agent}, \emph{service}, \emph{both}, or \emph{neither} as a recovery-oriented
classification.

The classification follows the available evidence. Without a valid target operation, the
injected service path leaves no task-specific trace, so any service contribution is
\emph{shadowed}. A valid timely operation followed by a missed state deadline supplies
service-side evidence under the study's fault model. Evidence of both shortfalls produces
\emph{both}; insufficient attribution evidence produces \emph{neither}, while unreadable
verifier evidence produces NotDetermined.

A late agent operation can overlap subsequent service convergence in the recorded trajectory.
The evaluation reports evidence coverage separately from conditional agreement with the
injected cause.

Attribution turns the verdict into a control decision. The completion gate releases a report
only after a Fulfilled verdict; recovery is not evidence of completion, so the resulting state
must be verified again. The classification selects the action: re-invoke the agent with the
canonical target after an agent-side shortfall, wait and re-observe after a service-side
shortfall, re-invoke and then wait after evidence of both, or escalate and perform one bounded
re-check when evidence is insufficient. Figure~\ref{fig:concept} summarizes this path from
contract instantiation to admitted completion or recovery.

\begin{figure}[h]
  \centering
  \resizebox{0.8\textwidth}{!}{\input{figures/fig0-concept-body}}
  \caption{Completion-contract architecture (red: agent, non-authoritative for completion;
  green: contract control plane;
  blue: distributed service). An operator instantiates a task contract from a template. The
  verifier observes service state outside the agent's context; its verdict gates completion,
  and the attribution informs recovery.}
  \label{fig:concept}
\end{figure}

Recovery does not erase a missed deadline. Re-invocation creates a new, identified attempt
over the same predicate and expected value, with a new verification window. A wait cycle keeps
the existing anchor and extends the observation deadline. In both cases, the earlier verdict
remains in the record. The record therefore distinguishes \emph{Fulfilled on the first
attempt}, \emph{Violated then verified after recovery}, and \emph{terminated unverified}.
Section~\ref{sec:design} specifies the cycle and wall-clock limits used in the evaluation.

The distinction becomes concrete when an agent targets the wrong service object. In the
evaluation, some models supplied a site's human-readable label where the service required an
internal identifier. The call was accepted and health remained green, but the target state did
not change. The verifier classified the missing valid target operation as agent-side; a new
attempt received the canonical identifier and was admitted only after the predicate was
observed.

The study instantiates this mechanism in a service with a bounded operation vocabulary and one
operator-defined template per task type. A natural-language goal becomes a contract through
template selection, parameter extraction, and operator confirmation. Scored runs start from
the resulting contract; the derivation pilot in Sect.~\ref{sec:supplementary} tests this process
on the study's templated goals. Deriving contracts from free-form goals is a separate problem;
the evaluation concerns what a contract does once it exists.

Whatever the template, the guarantee extends only as far as its predicate. A checksum or
side-effect constraint must appear in that predicate to be enforced. The evaluation addresses
fallible agents that misground or misgrade tasks; adversarial evasion is outside its threat
model.

%% file: figures/fig0-concept-body.tex
\begin{tikzpicture}[
  >=Stealth,
  font=\footnotesize,
  plane/.style={draw=green!55!black, fill=green!16, rounded corners=2pt, align=center,
                inner sep=3pt},
  agent/.style={draw=red!65!black, double, fill=red!14, rounded corners=2pt, align=center,
                inner sep=3pt},
  svc/.style={draw=blue!60!black, fill=blue!16, fill opacity=1, text opacity=1,
              rounded corners=2pt, align=center, inner sep=3pt},
  arr/.style={-{Stealth[length=2.2mm]}, semithick},
  ret/.style={-{Stealth[length=2.2mm]}, semithick, dashed},
  rem/.style={-{Stealth[length=2.4mm]}, thick, dashed, red!55!black},
  elbl/.style={font=\scriptsize, align=center, inner sep=1.5pt},
  trustframe/.style={draw=green!50!black, dashed, rounded corners=9pt, inner sep=7pt}
]
\node[plane, dashed, text width=2.5cm, minimum height=0.85cm] (op) at (0.0, 3.10)
  {\textbf{Operator}\\ (human / controller)};
\node[plane, text width=2.8cm, minimum height=1.3cm]  (ct) at (0.0, 0.35)
  {\textbf{Per-task contract}\\ condition, deadline,\\ expected value;\\ obligated party: agent};
\node[agent, text width=2.9cm, minimum height=1.05cm] (ag) at (3.5, 4.85)
  {\textbf{LLM agent}\\ authors operations,\\ reports ``done''};
\node[plane, text width=3.35cm, minimum height=1.5cm]  (vf) at (6.2, 0.60)
  {\textbf{Independent verifier}\\ polls state on a fixed grid\\ until the deadline $\Rightarrow$\\
   Fulfilled / Violated /\\ NotDetermined};
\node[plane, text width=2.7cm, minimum height=1.05cm] (gt) at (5.75, -2.17)
  {\textbf{Completion gate}\\ releases only on Fulfilled\\ (else: unverified)};
\node[plane, text width=2.9cm, minimum height=0.85cm] (at) at (0.0, -2.17)
  {\textbf{Attribution \& recovery}\\ agent / service\\ both / neither};

\node[trustframe, fit=(op)(ct)(vf)(gt)(at)] (plane) {};
\node[svc, text width=2.9cm, minimum height=1.05cm] (sv) at (8.7, 4.75)
  {\textbf{Distributed service}\\ state propagates with\\ delay (may lead or lag)};
\node[elbl, font=\scriptsize\itshape, text=green!45!black, anchor=north west, fill=white,
      inner sep=1pt] at ([yshift=-1pt]plane.south west) {contract control plane};

\draw[arr] (op) -- (ct)
  node[elbl, midway, right=2pt, fill=white, inner sep=1pt]{instantiates\\(from a template)};
\draw[arr] (op.north east) -- (ag.south west)
  node[elbl, pos=0.55, above left, fill=white, inner sep=1pt]{goal (as a\\person states it)};
\draw[arr] (ct.east) -- (vf.west)
  node[elbl, pos=0.34, above=3pt, fill=white, inner sep=0pt]
  {condition,\\deadline,\\expected value};
\draw[arr] (ag.north east) -- (sv.north west)
  node[elbl, pos=-0.20, above=10pt, fill=white, inner sep=1pt]{operations (tool calls)};
\draw[ret] (sv.west) -- (ag.east)
  node[elbl, midway, below]{read-backs:\\one site, ad hoc};
\draw[arr] (ag.south) -- (gt.north west)
  node[elbl, pos=0.80, left=1pt, fill=white, inner sep=1pt]{claims done};
\draw[arr] (vf.south) -- (gt.north east)
  node[elbl, pos=0.45, right=6pt, fill=white, inner sep=1pt]{verdict};
\draw[arr] (gt.west) -- (at.east)
  node[elbl, midway, above=2pt, fill=white, inner sep=1pt]{if not admitted};
\draw[rem] (at.west) -- (-1.95,-2.17) -- (-1.95,4.85) -- (ag.west);
\node[elbl, text=red!55!black, fill=white, inner sep=1pt, text width=3.4cm] at (0.0, 5.25)
  {recovery: re-invoke (agent-side) / wait (service-side)};
\draw[ret] (vf.east) to[bend right=14] (sv.south);
\node[elbl, fill=white, inner sep=1pt, text width=2.35cm, align=center] at (7.00, 2.70)
  {task-specific state probes,\\fixed grid; results\\bypass the agent};
\end{tikzpicture}

%% file: sections/04-evaluation-design.tex
\section{Evaluation Design}\label{sec:design}

The evaluation asks three questions. \textbf{RQ1:} How often do common completion policies
disagree with deadline-aware observation of service state, and in which direction?
\textbf{RQ2:} For what share of violations does runtime evidence support attribution between
agent and service, and how often does it agree with the injected cause when the required
evidence is observable? \textbf{RQ3:} What happens when the verdict is enforced as a
completion gate, including recovery outcomes and cost?

We study a live Onedata federation~\cite{onedata}: independently operated storage sites
exposed as one eventually consistent data-management service. Cross-site changes take time to
propagate even when calls succeed and components remain healthy. The study observes this
asynchrony in the live service rather than a simulated state machine.

Two task types exercise opposite state transitions (Table~\ref{tab:instances}): placement
establishes a replica at a named site, whereas deletion removes a file from every site. Their
predicates use target byte count for placement and path absence plus residual bytes for
deletion. Both use a 30\,s task deadline and 3\,s polling interval, fixed before the scored
runs.

\begin{table}[tb]
  \centering
  \caption{Evaluated contract instances. The predicates and expected values were fixed before
  each run; the agent was the obligated component in both tasks.}
  \label{tab:instances}
  \scriptsize
  \setlength{\tabcolsep}{4pt}
  \input{figures/tab1-instances-body}
\end{table}

We evaluate both tasks with seven models from five families: GLM-4.7-Flash and GLM-5.2-FP8;
Qwen3-Coder-30B; gemma-4-31B; claude-haiku-4-5 and claude-sonnet-5; and Llama-3.3-70B. Every
agent receives the same frozen task prompt, a five-tool MCP interface~\cite{mcp2025}, and eight
tool rounds. The interface exposes the reads needed to inspect task state and the task's single
mutating operation. Each full 56-trial cell contains eight runs per model, sampling
heterogeneous behavior across five model families.

Table~\ref{tab:matrix} maps the study cells to the RQs. Ungated records are re-scored for RQ1
and supply attribution evidence for RQ2; new gated executions, in which the verdict changes
runtime behavior, support RQ3. The injected fault was applied only to placement. Deletion
eligibility required setup to establish a converged two-site replica before the agent turn,
leaving 50 of 56 planned gated attempts. All other planned 56-trial cells were included in
full.

\begin{table}[tb]
  \centering
  \caption{Study cells used by the three RQs, reported as included/planned. A dash denotes a
  condition not run, not a zero outcome. Full cells hold seven models, eight runs each.\looseness=-1}
  \label{tab:matrix}
  \scriptsize
  \setlength{\tabcolsep}{5pt}
  \input{figures/tab-study-matrix-body}
\end{table}

The injected condition adds a 35\,s delay to the target's inter-provider transfer path. Because
the delay exceeds the 30\,s deadline, affected placement cannot meet the contract although
calls succeed and component health remains green. The condition isolates the evidence gap
between operational health and task fulfillment~\cite{basiri2016}. The injection record
establishes that the fault was active but is withheld from the runtime verifier. RQ2 counts
service-side evidence when a valid transfer exposes the injected path and marks the remaining
cases as shadowed.

Each ungated trial records the agent's tool-call trace and report, call outcomes, health
signals, timestamped verifier observations through and beyond the deadline, the instantiated
predicate, and injection status. Gated records also contain attempt identifiers, verdicts,
emitted attribution, recovery actions, and the final admitted or unverified outcome.

These records support five completion policies, ordered by what each can read:

\begin{enumerate}
\item \emph{Self-report}: accept if the agent says the task is complete.
\item \emph{Call-and-health monitoring}: accept unless a call or service-health alarm fires.
\item \emph{Transcript judge}: a fixed claude-sonnet-5 judge reads the agent's trace and final
answer, but not the service state; unparseable responses yield NotDetermined~\cite{advani2026}.\looseness=-1
\item \emph{State at report time}: read the service state once when the agent reports.
\item \emph{State until deadline}: observe the contracted predicate through the deadline
(Sect.~\ref{sec:contracts}).
\end{enumerate}

The fifth policy supplies the study's \emph{operational reference verdict}: its predicate,
deadline, and observations define whether a trial is Fulfilled. RQ1 measures how decisions
based on less or earlier evidence depart from that reference. Later observations of the
same predicate check the consistency of gate-admitted completions.\looseness=-1

The analysis distinguishes errors in both directions. A \emph{false pass} pairs policy
acceptance with a Violated reference verdict; a \emph{false alarm} pairs rejection or
NotDetermined with a Fulfilled reference verdict. Tables report trial-level incidence ($n/N$)
unless a caption gives a conditional denominator. Conditional false-pass and false-alarm rates
divide by the Violated and Fulfilled reference outcomes, respectively. Healthy placement
contains 23 Violated and 33 Fulfilled outcomes. NotDetermined remains a separate decision even
when grouped with non-acceptance for false-alarm incidence.

Enforcement adds task-level outcomes. In gated runs, an admitted report is a \emph{verified
completion}; it becomes a \emph{false completion} if later observations contradict the
admitted predicate. A predicate reached only after recovery is \emph{verified after recovery},
while the missed attempt remains Violated. A task that exhausts its cycle or wall-clock cap
without a Fulfilled attempt terminates \emph{unverified}. These classes preserve earlier
deadline violations. Later-contradiction rates divide by admitted completions; recovery and
termination counts use all gated trials.

The first verification cycle is anchored before the agent turn. Re-invocation starts a fresh
30\,s window; waiting extends the current window. Recovery is capped at three cycles and a
120\,s wall budget, checked between cycles.

We report aggregate counts across the evaluated panel. The panel varies agent behaviour; the
evaluation asks whether one completion criterion holds across that variation, not how the
models compare. Wilson 95\% intervals accompany the principal binomial proportions, including zero
counts.

Two supplementary analyses extend the design. A same-day control compared two 56-trial
healthy-placement arms: the original prompt and one that also supplied the contract's target,
predicate, deadline, and instruction to check until success or timeout. A second pilot applied
one fixed derivation prompt to the study's 56 placement and 56 deletion goals, then compared
the result with the operator-instantiated contract. Results report each analysis with its own
denominator. Recorded traces also provide the count of tool calls outside the task-specific
toolbox used in the admission-control analysis.

%% file: figures/tab1-instances-body.tex
\begin{tabular}{@{}>{\raggedright\arraybackslash}p{1.55cm}p{5.0cm}p{5.0cm}@{}}
\hline
 & \textbf{Placement task} & \textbf{Deletion task} \\
\hline
Goal &
``replicate this file to site B, and confirm it arrived'' &
``delete this file from everywhere --- not just hidden; confirm it is fully removed'' \\[1pt]
Fulfillment condition &
the file's byte count at the target site equals the write-time expected size &
the path is absent \emph{and} no site retains bytes for the pre-captured file identifier \\[1pt]
Verification &
state polled every 3\,s to the deadline; the service's own status signals are diagnostic only &
path-absence and per-site residue probes on the pre-captured identifier, polled to the
deadline \\[1pt]
Recovery &
routed by attribution: re-invoke with the canonical target (agent) $/$ wait (service) $/$ terminate as unverified &
same routing (Sect.~\ref{sec:enforcement}) \\
\hline
\end{tabular}

%% file: figures/tab-study-matrix-body.tex
\begin{tabular}{@{}l l c c l@{}}
\hline
\textbf{Mode} & \textbf{Task} & \textbf{Healthy I/P} & \textbf{Fault I/P} & \textbf{Primary use} \\
\hline
ungated & placement & 56/56 & 56/56 & RQ1; RQ2 \\
ungated & deletion & 56/56 & --- & RQ1; organic violations \\
gated & placement & 56/56 & 56/56 & RQ3 \\
gated & deletion & 50/56 & --- & RQ3 \\
\hline
\end{tabular}

%% file: sections/05-results.tex
\section{Results}\label{sec:results}

\subsection{Completion-policy errors (RQ1)}\label{sec:checksbreak}

In the 56 healthy placement trials, the deadline-aware reference classified 33 as Fulfilled
and 23 as Violated. Self-report erred in both directions: agents accepted 21 violations and
made no completion report for nine fulfilled trials. The reference returned no NotDetermined
verdict in the healthy placement, faulted placement, or deletion populations.

Twenty of those 21 false reports followed at least one agent read-back. The recorded signals
included contrary target state, successful job submission, a directory aggregate, and
convergence after the deadline (Table~\ref{tab:readback}). Every false deletion report likewise
followed a post-delete read.

\begin{table}[t]
  \caption{Recorded read-back preceding each of the 21 false completion reports in the
  healthy ungated placement cell. Counts sum to 21.}
  \label{tab:readback}
  \centering
  \scriptsize
  \input{figures/tab2-readback-classes-body}
\end{table}

Table~\ref{tab:misjudgment} reports error incidence over all 56 trials in each placement cell.
Call-and-health monitoring accepted 14 of 23 healthy violations and 47 of 56 faulted
violations. Under the injected delay, successful calls and green health did not establish
arrival at the target. The transcript judge accepted none of the 23 healthy violations but
failed to accept 20 of 33 fulfilled trials. Eight outputs were unparseable and therefore
NotDetermined, split evenly between the two reference classes. A report-time state read made
one false pass.

\begin{table}[tb]
  \centering
  \caption{Placement-policy error incidence, shown as $n/56$ (percentage) within each cell.
  Healthy placement contains 23 Violated and 33 Fulfilled reference outcomes; all 56 injected-
  fault outcomes are Violated.}
  \label{tab:misjudgment}
  \scriptsize
  \input{figures/tab3-placement-misjudgment-body}
\end{table}

Table~\ref{tab:deadline} re-scores the same trajectories at different deadlines. At 30\,s, four
of the 23 healthy violations later converged; the other 19, caused by absent or incorrectly
targeted operations, remained Violated throughout the observation horizon. All injected-delay
trajectories remained Violated through 40\,s, and no policy comparison changed direction from
25 to 40\,s.

\begin{table}[tb]
  \centering
  \caption{Deadline sensitivity from re-scoring the same 56 healthy and 56 faulted placement
  trajectories at each $T$. “Healthy late cases” counts trajectories observed to converge
  within the recorded horizon, but only after $T$; the 19 never-convergent healthy trajectories
  remain Violated but are not in that row. “Fault passes” counts faulted trajectories satisfied
  by $T$.}
  \label{tab:deadline}
  \scriptsize
  \input{figures/tab10-deadline-sensitivity-body}
\end{table}

Deletion produced a different pattern among 16 Violated and 40 Fulfilled reference outcomes.
Self-report, call-and-health monitoring, and the report-time read accepted 12, 11, and 9 of the
16 violations, respectively (Table~\ref{tab:deletion}). The transcript judge accepted no
violation but also no fulfilled deletion. Its 56 verdicts comprised 44 NotDetermined (35 on
Fulfilled and nine on Violated outcomes) and 12 Violated.

\begin{table}[tb]
  \centering
  \caption{Deletion false passes as trial incidence and conditional on the 16 Violated
  among 56 ungated trials. The transcript judge also false-alarmed on all 40
  Fulfilled.\looseness=-1}
  \label{tab:deletion}
  \scriptsize
  \input{figures/tab5-deletion-misjudgment-body}
\end{table}

\subsection{Evidence coverage for attribution (RQ2)}\label{sec:attribution}

The injected delay was active in all 56 faulted placement trials. One operation lay outside
the injection window and had ground-truth class \emph{neither}, leaving 55 trials
ground-truthable against the injected cause. In 18, no valid target operation exposed the
service path, so its contribution was shadowed. All 37 trials with the required operation and
state evidence agreed with the injected cause (95\% Wilson lower bound 90.6\%). Evidence
coverage was therefore 37/55 (67\%), with conditional agreement of 37/37 (100\%; Table~\ref{tab:attribution}).

\begin{table}[tb]
  \centering
  \caption{Attribution accounting for the 56 injected-fault placement trials. One
  truth-\emph{neither} trial lies outside the injected-cause denominator; shadowed trials
  remain in the ground-truthable denominator.}
  \label{tab:attribution}
  \scriptsize
  \input{figures/tab6-attribution-body}
\end{table}

As a secondary check, blinded author adjudication of the 39 violations without an injected
fault agreed with the emitted attribution in 28 cases (72\%). The 11 disagreements concentrated
in failed control-API calls and operations issued too late to separate agent delay from
subsequent convergence.

\subsection{Enforced completion and recovery (RQ3)}\label{sec:enforcement}

Across 112 gated placement trials, 109 ended in an admitted completion and three terminated
unverified. Later observations contradicted none of the 109 admitted completions (95\% Wilson
upper bound 3.4\%). All 56 healthy trials reached verified completion. Under the injected
delay, every first attempt missed its deadline; 53 trials recovered and three terminated
unverified (Table~\ref{tab:gateoutcomes}). In all 53 records the first attempt keeps its
verdict.

The 50 included healthy gated deletion trials also all reached verified completion, with no
later contradiction (0/50; upper bound 7.1\%).

One placement trial ended with a NotDetermined verdict and no deletion trial did.
NotDetermined is a verifier verdict, whereas \emph{unverified} is task-level termination after
the recovery budget; only one of the three unverified trials ended NotDetermined.

\begin{table}[tb]
  \centering
  \caption{Outcomes of the 112 gated placement trials. “Recovered” means the first attempt
  failed and a later attempt was verified; it does not revise the first verdict.}
  \label{tab:gateoutcomes}
  \scriptsize
  \input{figures/tab9-gate-outcomes-body}
\end{table}

Healthy gated placement recovered 29 failed first attempts, 16 through re-invocation and 13
through the escalation re-check (Table~\ref{tab:rescue}).

Fixed recovery policies can be bounded on these records. Waiting cannot create a missing
operation, so an always-wait policy completes at most 40 of 56 healthy and 39 of 56 faulted
trials, against 56 and 53 under routed recovery. An always-re-invoke policy would re-engage the
agent in 13 healthy and 39 faulted trials that routing completed without it.

\begin{table}[tb]
  \centering
  \caption{First-attempt failures recovered by mechanism in the 56 healthy gated placement
  trials. Each model contributes eight trials.}
  \label{tab:rescue}
  \scriptsize
  \input{figures/tab7-rescue-mechanism-body}
\end{table}

Table~\ref{tab:cost} describes recovery activity with an unweighted index: the sum of verifier
probes, agent invocations, and gate retries divided by verified completions. It is neither
elapsed time nor monetary cost, and its per-model values are descriptive rather than a
ranking.\looseness=-1

\begin{table}[tb]
  \centering
  \caption{Unweighted runtime-activity index per verified placement completion. Each entry sums
  verifier probes, agent invocations, and gate retries, divides by the number of verified
  completions, and gives the three event types equal weight.}
  \label{tab:cost}
  \scriptsize
  \input{figures/tab8-cost-body}
\end{table}

Direct billing isolates the gate's cost for the two SDK-billed models, each run with and
without the gate (Table~\ref{tab:dollars}). Gating added 5--17\% to model spend. Under the
fault, both models reached the predicate in 8 of 8 gated trials, after recovery, and in 0 of 8
ungated trials; the gate does not make faulted tasks meet the deadline. Charges use the SDK's July 2026 billed token accounting, including tool-use tokens.

The verifier issued about 32 probes per placement trial without the fault, gated or not, and
about 56 per faulted gated trial, where recovery windows add probes.
A separate benchmark with up to 100 open contracts measured the verifier's own cost: at the
3\,s interval, about 0.2 CPU core-seconds per contract per minute (100 contracts: a third of one
core, 54\,MiB), no probe failures, and no change in another client's median latency. The cost
followed the polling rate, from 0.13 at 5\,s to 0.44--0.59 core-seconds at 1\,s.

\begin{table}[tb]
  \centering
  \caption{Recorded model charges (USD) in eight-trial cells for the two SDK-billed models, with
  and without the gate. Ungated counts completions within the deadline; gated counts trials
  that reached the predicate, including after recovery. Service and infrastructure costs are
  excluded: the federation is shared, and its cost cannot be reliably attributed to individual
  tasks; the verifier's cost is reported as its probe count.\looseness=-1}
  \label{tab:dollars}
  \scriptsize
  \input{figures/tab11-cost-dollars-body}
\end{table}

\subsection{Supplementary analyses}\label{sec:supplementary}

The first supplementary analysis asks whether giving the contract to the agent improves its
actions. A post hoc, same-day paired control compared the original prompt with one that also
supplied the condition, deadline, and instruction to check until success or timeout (56 trials
per arm). Providing the exact target eliminated wrong-target schedules (10/56 to 0), reduced
omitted transfers (11/56 to 5/56), and reduced reference violations from 23/56 to 8/56.
Contract information improved action selection without making self-report authoritative: the
uninformed arm reported all 23 violations complete, and the informed arm reported six of
eight complete.

Before the completion decision, admission control blocked all 117 attempted calls outside the
task-specific toolbox, including 64 to a rule-setting operation that offered another placement
path. In a separate derivation pilot, a fixed prompt reconstructed the instantiated contract
for all 56 placement and all 56 deletion goals (95\% Wilson lower bound 93.6\% for each task
family). The pilot measured consistency on the study's templated goals.

%% file: figures/tab2-readback-classes-body.tex
\begingroup\setlength{\tabcolsep}{4pt}
\begin{tabular}{@{}p{3.4cm}c p{6.0cm}@{}}
\hline
\textbf{Read-back class} & \textbf{$n$ (of 21)} & \textbf{What the recorded read showed} \\
\hline
Read, then ignored & 6 & Nothing at the target; reported done anyway. \\[1pt]
Job status, not data & 5 & Took the transfer job's successful submission as arrival; the target itself still held zero bytes. \\[1pt]
Late-but-real convergence & 5 & Placement genuinely converged, but only past the 30-second deadline. \\[1pt]
Wrong signal (directory aggregate) & 4 & Read a nonzero parent-directory total, mistaking it for the target file. \\[1pt]
No read-back & 1 & No read-back issued before reporting done. \\
\hline
\end{tabular}
\endgroup

%% file: figures/tab3-placement-misjudgment-body.tex
\begingroup\setlength{\tabcolsep}{5pt}
\begin{tabular}{@{}p{4.4cm}ccc@{}}
\hline
\textbf{Completion policy} & \textbf{Healthy FP} & \textbf{Healthy FA} & \textbf{Fault FP} \\
\hline
self-report & 21/56 (38\%) & 9/56 (16\%) & 37/56 (66\%) \\[1pt]
call/health monitoring & 14/56 (25\%) & 1/56 (2\%) & 47/56 (84\%) \\[1pt]
transcript judge & 0/56 & 20/56 (36\%) & 1/56 (2\%) \\[1pt]
state at report time & 1/56 (2\%) & 0/56 & 0/56 \\[1pt]
state until deadline (reference) & 0/56 & 0/56 & 0/56 \\
\hline
\end{tabular}
\endgroup

%% file: figures/tab10-deadline-sensitivity-body.tex
\begin{tabular}{@{}l c c c c c c@{}}
\hline
\textbf{Deadline $T$ (s)} & \textbf{10} & \textbf{20} & \textbf{25} & \textbf{30 (used)} & \textbf{35} & \textbf{40} \\
\hline
healthy late cases at $T$ & 37 & 22 & 13 & 4 & 2 & 1 \\[1pt]
fault passes & 0 & 0 & 0 & 0 & 0 & 0 \\
\hline
\end{tabular}

%% file: figures/tab5-deletion-misjudgment-body.tex
\begingroup\setlength{\tabcolsep}{6pt}
\begin{tabular}{@{}p{5.2cm}cc@{}}
\hline
\textbf{Completion policy} & \textbf{FP incidence} & \textbf{FP among Violated} \\
\hline
self-report & 12/56 (21\%) & 12/16 (75\%) \\[1pt]
call/health monitoring & 11/56 (20\%) & 11/16 (69\%) \\[1pt]
transcript judge & 0/56 & 0/16 \\[1pt]
state at report time & 9/56 (16\%) & 9/16 (56\%) \\[1pt]
state until deadline (reference) & 0/56 & 0/16 \\
\hline
\end{tabular}
\endgroup

%% file: figures/tab6-attribution-body.tex
\begingroup\setlength{\tabcolsep}{6pt}
\begin{tabular}{@{}l c@{}}
\hline
\textbf{Attribution quantity} & \textbf{Observed result} \\
\hline
ground-truthable against injected cause & 55/56 (98\%) \\
evidence observable & 37/55 (67\%) \\
shadowed by missing valid operation & 18/55 (33\%) \\
agreement when evidence observable & 37/37 (100\%) \\
\hline
\end{tabular}
\endgroup

%% file: figures/tab9-gate-outcomes-body.tex
\begin{tabular}{@{}l c c@{}}
\hline
\textbf{Gate outcome} & \textbf{Healthy (56)} & \textbf{Fault (56)} \\
\hline
verified completions & 56 & 53 \\
later-contradicted completions & 0 & 0 \\
unverified terminations (explicit) & 0 & 3 \\
trials recovered after a failed first attempt & 29 (52\%) & 53 (95\%) \\
runtime-activity index per verified completion (mean) & 17.88 & 32.58 \\
\hline
\end{tabular}

%% file: figures/tab7-rescue-mechanism-body.tex
\begin{tabular}{@{}l c c c@{}}
\hline
\textbf{Model} & \textbf{Rescued (/8)} & \textbf{By re-invoke} & \textbf{By re-check} \\
\hline
Qwen3-Coder-30B & 6 & 6 & 0 \\
Llama-3.3-70B & 5 & 5 & 0 \\
GLM-4.7-Flash & 4 & 4 & 0 \\
GLM-5.2-FP8 & 4 & 0 & 4 \\
claude-sonnet-5 & 8 & 1 & 7 \\
claude-haiku-4-5 & 2 & 0 & 2 \\
gemma-4-31B & 0 & 0 & 0 \\
\hline
\textbf{total} & \textbf{29} & \textbf{16} & \textbf{13} \\
\hline
\end{tabular}

%% file: figures/tab8-cost-body.tex
\begin{tabular}{@{}p{2.6cm}>{\centering\arraybackslash}p{2.3cm}>{\centering\arraybackslash}p{2.3cm}@{}}
\hline
& \multicolumn{2}{c}{\textbf{Runtime-activity index}} \\
\textbf{Model} & \textbf{Healthy} & \textbf{Injected delay} \\
\hline
gemma-4-31B & 9.12 & 28.00 \\
claude-haiku-4-5 & 11.88 & 24.00 \\
GLM-5.2-FP8 & 15.50 & 31.71 \\
Llama-3.3-70B & 19.25 & 50.50 \\
claude-sonnet-5 & 20.50 & 20.12 \\
GLM-4.7-Flash & 20.50 & 36.62 \\
Qwen3-Coder-30B & 28.38 & 41.50 \\
\hline
\end{tabular}

%% file: figures/tab11-cost-dollars-body.tex
\setlength{\tabcolsep}{6pt}\begin{tabular}{@{}l l r r r r r@{}}
\hline
 & & \multicolumn{2}{c}{\textbf{Ungated}} & \multicolumn{2}{c}{\textbf{Gated}} & \\
\textbf{Model} & \textbf{Condition} & in time & USD & reached & USD & \textbf{Spend change} \\
\hline
claude-haiku-4-5 & healthy & 8$/$8 & 0.2532 & 8$/$8 & 0.2826 & +11.6\% \\
 & fault & 0$/$8 & 0.3343 & 8$/$8 & 0.3518 & +5.2\% \\
claude-sonnet-5 & healthy & 5$/$8 & 1.5733 & 8$/$8 & 1.8380 & +16.8\% \\
 & fault & 0$/$8 & 1.8811 & 8$/$8 & 2.0627 & +9.7\% \\
\hline
\end{tabular}

%% file: sections/06-discussion.tex
\section{Discussion}\label{sec:discussion}

The results support separating execution from completion authority. The agent selects and
executes operations; a control-plane component with a pre-specified predicate, a clock, and
independent observations decides whether the resulting service state satisfies the task. As in
other service-control mechanisms, the component doing work emits evidence, while a separate
mechanism decides whether an objective was met.

The RQ1 comparisons show why this distinction matters. Self-report, health monitoring, transcript
judging, and report-time state reads decide from evidence available at different times and
scopes. A deadline verdict instead asks whether the contracted state became observable within
the specified window. Supplying the contract to the agent improved operation choices in the
supplementary control, while independent admission still supplied the completion decision.
The two interventions are complementary.\looseness=-1

This separation also clarifies the heterogeneous-model result. In the evaluated healthy
placement cell, models differed in first-attempt success and recovery activity, while the gate
applied one completion criterion to all of them. Model capability determines how the predicate
is reached; the gate determines when completion is admitted. A system can therefore expose
model differences as latency, retries, or cost without changing the stated meaning of
``complete.''

Separating authority does not by itself determine the strength of the guarantee; the
predicate sets that boundary. Byte counts and
path residue are inexpensive service-state checks; checksums, atomic snapshots, and side-effect
constraints provide stronger assurance but demand richer observability. The contract exposes
this trade-off at task definition time.

The predicate governs outcomes, whereas tool admission governs which actions may occur.
Restricting the offered toolbox prevented
off-tool calls from executing, and the completion gate controlled whether
the resulting task state could be reported as done. A deployable agent-service architecture
needs both: action-time controls for operations that must never execute and completion-time
controls for outcomes that must be observed before release.

When the predicate is missed, the control plane must also decide how to recover. That decision
depends on evidence from both agent action and service state. A missing target
operation can shadow a service contribution, while a late operation can overlap subsequent
service convergence. The runtime used the emitted classification to select re-invocation and
wait recoveries. The completion gate is the primary result; attribution is a secondary mechanism,
validated where its evidence was observable (37 of 55 cases). Its errors can slow recovery but
cannot admit a false completion. A production interface should make \emph{not attributable} a first-class
outcome for such ambiguous trajectories.

Beyond a single task, per-task verdicts could become inputs to service-level aggregates such as
error budgets, and a multi-stage workflow could require a Fulfilled stage verdict before the
next stage begins. Composition raises additional questions about deadline allocation,
compensation for in-flight effects, and recovery across stage boundaries. The current
single-task contract provides the unit from which these workflow guarantees can be composed.

%% file: sections/07-threats.tex
\section{Threats to Validity}\label{sec:threats}

\paragraph{Construct validity.}

The operational reference captures only the contracted predicate within the 30\,s deadline.
Placement checks byte count, not contents; deletion checks path absence and per-site residue,
not an atomic federation snapshot. The runtime-activity index measures neither time nor
monetary cost. Attribution reports evidence coverage separately from conditional agreement;
the organic comparison relies on blinded secondary author adjudication.

\paragraph{Internal validity.}

The trials ran on a live shared federation, where background load, convergence, run timing and
order, and model endpoints may affect rates. Fixed prompts define one agent configuration. The
35\,s injected delay exceeds the 30\,s deadline and represents one fault mechanism.

\paragraph{External validity.}

The study covers one live federated data service, two tasks, seven models, and one injected
fault class, applied only to placement. Other services may require different predicates and
observations. Deployment assumes timely read-only access to authoritative service state and
stable task identifiers.

%% file: sections/08-conclusion.tex
\section{Conclusion}\label{sec:conclusion}

When an LLM agent consumes a service, execution and completion authority need not reside in
the same component. The agent may choose operations and inspect intermediate state, while an
independent control plane decides whether a pre-specified task predicate held within its time
window. This separation is the paper's main service-computing contribution. It turns
``done'' from a conversational claim into an observable runtime decision with an explicit
outcome and recovery path.

The live evaluation shows what that separation changes. Completion policies based on agent
reports, health signals, transcripts, or one state read diverged from deadline-aware service
state in different directions. Under enforcement, the same admission rule applied across a
heterogeneous model panel: no admitted completion was later contradicted for the evaluated
predicates, healthy tasks completed, and unresolved faulted tasks terminated explicitly
rather than silently passing. Model behavior still affected retries, latency, and cost. It no
longer determined the semantics by which completion was admitted.

The evaluation makes consumer-obligated SLA semantics operational at task granularity: specify
the required
state before execution, observe it independently, preserve deadline violations through
recovery, and make the verdict authoritative at the completion boundary. The selected
predicate defines what completion means, while observable agent and service evidence directs
recovery.

Future work should first validate contract derivation on goals and services beyond the two
templates used here and make NotDetermined and not-attributable outcomes explicit in the
runtime interface. A second priority is a controlled comparison of attribution-based recovery
with always-wait and always-re-invoke policies. Broader evaluations should then cover other
service classes, gentler and mixed faults, and multi-stage contracts whose deadlines and
recoveries compose across a workflow.

%% file: artifact/fig-setup-body.tex
\begin{tikzpicture}[font=\scriptsize,
  box/.style={draw, rounded corners=2pt, minimum width=1.8cm, minimum height=0.6cm, text width=1.7cm, align=center},
  svc/.style={draw, rounded corners=2pt, fill=gray!8, minimum width=3.4cm, text width=3.3cm, align=center},
  arr/.style={-{Stealth[length=4pt]}}]
  \node[box] (agent) at (1.3,2.75) {Agent (LLM)};
  \node[box] (gate)  at (1.3,1.55) {Gate};
  \node[box] (mcp)   at (4.2,2.75) {onedata-mcp};
  \node[box] (ver)   at (4.2,1.55) {Verifier};
  \node[draw, dashed, rounded corners=3pt, fit=(agent)(gate)(mcp)(ver), inner sep=5pt,
        label={[anchor=north west, inner sep=1pt]south west:Harness host}] (host) {};
  \node[box, minimum width=2.6cm, text width=2.5cm] (models) at (1.3,4.2) {Model endpoints};
  \node[svc, minimum height=0.75cm] (oz)  at (8.6,3.75) {Onezone\\Helm chart \texttt{onezone}};
  \node[svc, minimum height=1.6cm]  (src) at (8.6,2.15) {Oneprovider (source site)\\Helm chart \texttt{oneprovider}};
  \node[svc, minimum height=0.75cm] (tgt) at (8.6,0.35) {Oneprovider (target site)\\Helm chart \texttt{oneprovider}};
  \draw[arr] (agent) -- (models) node[midway, right] {LLM API};
  \draw[arr] (agent) -- (mcp) node[midway, above] {tools};
  \draw[arr] (agent) -- (gate) node[midway, left] {report};
  \draw[arr] (ver) -- (gate) node[midway, above] {verdict};
  \draw[arr] (mcp.east) -- (mcp.east -| src.west) node[midway, above] {REST};
  \draw[arr] (ver.east) -- (ver.east -| src.west) node[midway, above] {state reads};
  \draw[dashed] (oz) -- (src) node[midway, right] {federates};
  \draw[arr] (src.east) -- (10.6,2.15) |- (tgt.east) node[pos=0.25, right, text width=1.1cm, align=left] {transfer\\(delay\\injected)};
\end{tikzpicture}